\documentclass{aa}
\usepackage[varg]{txfonts}
\usepackage{longtable}
\usepackage{color}

\usepackage{natbib,twoopt}
\usepackage[breaklinks=true, draft]{hyperref} 
\bibpunct{(}{)}{;}{a}{}{,}             

\makeatletter
  \newcommandtwoopt{\citeads}[3][][]{\href{http://adsabs.harvard.edu/abs/#3}%
    {\def\hyper@linkstart##1##2{}%
     \let\hyper@linkend\@empty\citealp[#1][#2]{#3}}}
  \newcommandtwoopt{\citepads}[3][][]{\href{http://adsabs.harvard.edu/abs/#3}%
    {\def\hyper@linkstart##1##2{}%
     \let\hyper@linkend\@empty\citep[#1][#2]{#3}}}
  \newcommandtwoopt{\citetads}[3][][]{\href{http://adsabs.harvard.edu/abs/#3}%
    {\def\hyper@linkstart##1##2{}%
     \let\hyper@linkend\@empty\citet[#1][#2]{#3}}}
  \newcommandtwoopt{\citeyearads}[3][][]%
    {\href{http://adsabs.harvard.edu/abs/#3}
    {\def\hyper@linkstart##1##2{}%
     \let\hyper@linkend\@empty\citeyear[#1][#2]{#3}}}
\makeatother
\newcommand\kms{$\mbox{km s}^{-1}$}

\newcommand\p{$\phantom{^{a}}$}

\newcommand\z{$\phantom{0}$}
\makeatletter
\renewcommand*\aa@pageof{, page \thepage{} of \pageref*{LastPage}}
\makeatother

\begin{document}

\title{The VLT-FLAMES Tarantula Survey}

\subtitle{XXXI. Radial velocities and multiplicity constraints of red supergiant stars in 30 Doradus\thanks{Based on observations collected at the European Organisation for Astronomical Research in the Southern Hemisphere under ESO programme 182.D-0222}
}

\author{L. R. Patrick\inst{1, 2}
  \and D. J. Lennon\inst{3, 1}
  \and N. Britavskiy\inst{1, 2}
  \and C. J. Evans\inst{4}
  \and H. Sana\inst{5}
  \and W. D. Taylor\inst{4}
  \and A. Herrero\inst{1, 2}
  \and L.~A.~Almeida\inst{6}
  \and J.~S. Clark\inst{7}
  \and M. Gieles\inst{8, 9, 10}
  \and N. Langer\inst{11}
  \and F. R. N. Schneider\inst{12}
  \and J. Th. van Loon\inst{13}
   }

\offprints{lpatrick@iac.es}
\authorrunning{L. R. Patrick et al.}  
\titlerunning{Radial velocities and multiplicity constraints of RSGs in 30~Doradus}

\institute{Instituto de Astrof\'isica de Canarias, E-38205 La Laguna, Tenerife, Spain \email{lpatrick@iac.es}
  \and Universidad de La Laguna, Dpto. Astrof\'isica, E-38206 La Laguna, Tenerife, Spain
  \and Department of Computer Science and Mathematics,
European Space Astronomy Centre (ESAC), Camino bajo del Castillo s/n, Urbanizacion Villafranca del Castillo,
Villanueva de la Cañada, 28 692 Madrid, Spain
\and UK Astronomy Technology Centre, Royal Observatory Edinburgh, Blackford Hill, Edinburgh, EH9 3HJ, UK
\and Institute of astrophysics, KU Leuven, Celestijnlaan 200D, 3001, Leuven, Belgium
\and Departamento de F\'isica Te\'orica e Experimental, Universidade Federal do Rio Grande do Norte, CP 1641, Natal, RN, Brazil
\and School of Physical Sciences, The Open University, Walton Hall, Milton Keynes, MK7 6AA, UK
\and Department of Physics, University of Surrey, Guildford GU2 7XH, UK
\and Institut de Ci\`{e}ncies del Cosmos (ICCUB), Universitat de Barcelona, Mart\'{i} i Franqu\`{e}s 1, E08028 Barcelona, Spain
\and ICREA, Pg. Lluis Companys 23, 08010 Barcelona, Spain.
\and Argelander-Institüt für Astronomie, Universität Bonn, Auf dem Hügel 71, D-53121 Bonn, Germany
\and Department of Physics, University of Oxford, Keble Road, Oxford OX13RH, UK
\and Lennard-Jones Laboratories, Keele University, Staffordshire ST5 5BG, UK
}

\date{Received ... / Accepted ...}

\abstract{}
{The incidence of multiplicity in cool, luminous massive stars is relatively unknown compared to their hotter counterparts. 
Here we present radial velocity (RV) measurements and investigate the multiplicity properties of red supergiants (RSGs) in the 30~Doradus region of the Large Magellanic Cloud using multi-epoch visible spectroscopy from the VLT-FLAMES Tarantula Survey.}
{Exploiting the high density of absorption features in visible spectra of cool stars, we use a novel slicing technique to estimate RVs of 17 candidate RSGs in 30~Doradus from cross-correlation of the observations with model spectra.}
{We provide absolute RV measurements (precise to better than $\pm$1\,\kms) for our sample and estimate line-of-sight velocities for the Hodge\,301 and SL\,639 clusters, which agree well with those of hot stars in the same clusters. By combining
results for the RSGs with those for nearby B-type stars, we estimate systemic velocities and line-of-sight velocity dispersions for the two clusters, obtaining estimates for their dynamical masses of $\log (M_{\rm dyn}/M_{\odot})=$~3.8\,$\pm$\,0.3 for Hodge\,301, and an upper limit of $\log (M_{\rm dyn}/M_{\odot})<$~3.1\,$\pm$\,0.8 for SL\,639, assuming Virial equilibrium.
Analysis of the multi-epoch data reveals one RV-variable, potential binary candidate (VFTS\,744), which is likely a semi-regular variable asymptotic giant branch star.
Calculations of semi-amplitude velocities for a range of RSGs in model binary systems and literature examples of binary RSGs were used to guide our RV variability criteria. We estimate an upper limit on the observed binary fraction for our sample of 0.3, where we are sensitive to maximum periods for individual objects in the range of 1 to 10\,000 days and mass-ratios above 0.3 depending on the data quality.
From simulations of the RV measurements from binary systems given the current data we conclude that systems within the parameter range q~$>$~0.3, $\log$P\,[days]~$<$~3.5, would be detected by our variability criteria, at the 90\% confidence level. 
The intrinsic binary fraction, accounting for observational biases, is estimated using simulations of binary systems with an empirically defined distribution of parameters where orbital periods are uniformly distributed in the 3.3~$<\log$P\,[days]~$<$4.3 range. A range of intrinsic binary fractions are considered; a binary fraction of 0.3 is found to best reproduce the observed data.}
{We demonstrate that RSGs are effective extragalactic kinematic tracers by estimating the kinematic properties, including the dynamical masses of two LMC young massive clusters.
In the context of binary evolution models, we conclude that the large majority of our sample consists of currently effectively single stars
(either single or in long period systems).
Further observations at greater spectral resolution and/or over a longer baseline are required to search for such systems.}

\maketitle

\section{Introduction}     \label{sec:introduction}

A clear picture has emerged that most massive stars reside in multiple systems
\citep{2007ARA&A..45..481Z,2009AJ....137.3358M,2011IAUS..272..474S,2012MNRAS.424.1925C,2014ApJS..215...15S}, with $\sim$70\% expected to interact with a companion during their lifetimes 
\citep[e.g.][]{2012Sci...337..444S,2013A&A...550A.107S,2014ApJS..215...15S,2013ARA&A..51..269D,2014ApJS..213...34K,2015A&A...580A..93D,2017ApJS..230...15M}. The ramifications of this are now being explored in 
detailed simulations of binary populations, to understand the impact on, e.g., massive-star evolution~\citep{2008MNRAS.384.1109E,2013MNRAS.436..774E,2013ApJ...764..166D,2017PASA...34....1D} and the timescales of
core-collapse supernovae~\citep{1992ApJ...391..246P,2010ApJ...725..940Y,2017A&A...601A..29Z}.

In this context, the cool part of massive-star evolution has received relatively little attention. To some extent the assumption has been that as a massive star cools (extending its radius
significantly) any close binary systems will simply interact and/or
merge, such that the star may not even make it all the way across the Hertzsprung--Russell (H--R) diagram, i.e., only 
currently single (or effectively single) stars will evolve to sufficiently large radii so as to become red supergiants (RSGs).
Indeed, the large radial extent of RSG atmospheres limits the potential companions in a binary
system compared to hotter stars. The most reliable observational estimates of RSG radii are of order  650--1500\,R$_{\odot}$~\citep{Wittowski12,2017A&A...606L...1W,2013A&A...554A..76A,2015A&A...575A..50A}, which immediately rule out short-period systems of the type found in hot stars~\citep[e.g.][]{2012Sci...337..444S,2013A&A...550A.107S,2017A&A...598A..84A,2018arXiv180405607L}.
Nonetheless, placing firm constraints on multiplicity in the RSG phase is important, both in terms of understanding the pathways for binary and single stellar evolution, as the progenitors of type II supernovae and particularly relevant for stripped-envelope supernovae of types IIb and Ib,c ~\citep[e.g.][]{Smartt09,2013MNRAS.436..774E,2018A&A...615A..78G}. 

In contrast to efforts for OB-type stars, multi-epoch studies of the radial velocities (RVs) of RSGs have been somewhat limited to date~\citep[e.g.][]{1928MNRAS..88..660S,1933ApJ....77..110S,1989AJ.....98.2233S} and the campaign by \citet{2007A&A...469..671J} provides a useful reference sample. They used 
high-resolution ($R$\,$\sim$\,40\,000) spectroscopy of 13 Galactic RSGs to estimate precise RVs (to better than 0.1\,\kms) over a 15 month period. These authors estimated `atmospheric' RVs for each target at each epoch, finding two groups -- the first with low-level variations ($\delta$RV\,$<$\,5\,\kms) and the second with variations of up to $\sim$10\,\kms\/ over the full baseline of their observations (see their Fig.~3)\footnote{Although note that four spectroscopic binaries (M-type supergiants with B-type companions) in their initial sample of 23 targets were not followed-up by their multi-epoch observations.}. In their characterisation of the velocities in different atmospheric layers of their targets, they attributed the variable velocity fields as probably originating from atmospheric convective cells, which is supported by theoretical studies that predict RV variations on similar scales \citep{1975ApJ...195..137S,2002AN....323..213F,2010ApJ...725.1170S}.

Examples of known RSGs in binary systems can also help guide expectations. VV~Cep-type binaries are eclipsing binary systems that have generally been identified by peculiar, variable spectra arising from a companion (typically a B-type star) shining through the atmosphere of the RSG.
Some $\zeta$~Aur-type binaries contain K-type supergiants with B-type secondaries and are by definition eclipsing systems~\citep{2017ars..book.....L}\footnote{However, although the $\zeta$~Aur stars are typically assumed to be supergiants, estimates of their stellar parameters suggest that the majority are lower-mass objects~\citep[e.g.][]{1996ApJ...471..454B,2007AJ....133.2669E}.}.
Typical semi-amplitude velocities within these systems are in the range of 20 to 30\,\kms~\citep[e.g.][]{1977JRASC..71..152W, 2007AJ....133.2669E}.

Motivated to obtain better empirical constraints on multiplicity
in the cool part of the H--R diagram, we have
turned to the cool stars observed as part of the VLT-FLAMES Tarantula Survey~\citep[VFTS;][hereafter Paper~I]{2011A&A...530A.108E}. The VFTS obtained multi-epoch optical spectroscopy of $\sim$1000 stars in the 30~Doradus region of the Large Magellanic Cloud (LMC).
Although primarily focused on the OB-type population in this region, the unbiased target selection led to the inclusion of
91 later-type objects (Table~3 in Paper~I).

The VFTS observations have been used to estimate the binary fraction of O- and B-type stars \citep[][respectively]{2013A&A...550A.107S,2015A&A...580A..93D}. Here we present a similar RV analysis (albeit using very different methods and sample sizes) of the K- and M-type stars observed in the VFTS to investigate their multiplicity. A parallel study (Britavskiy et al. submitted; hereafter B19) presents stellar parameters for the same sample and investigates their evolutionary status. 

This paper is organised as follows. In Sect.~\ref{sec:observations} we describe the observations and target selection for the study.
Sect.~\ref{sec:rv} presents the RV estimates and a comparison of our results to literature measurements.
In Sect.~\ref{sec:Mdyn} we estimate the kinematic properties and dynamical masses of the two lower-mass clusters in the 30~Dor region, and highlight the effectiveness of RSGs as kinematic tracers.
Sect.~\ref{sec:bin_anal} presents a first look at a binarity analysis of our sample using the multi-epoch data where we determine an upper limit to the observed binary fraction and the intrinsic binary fraction of our sample using simulations.
The key conclusions are presented in Sect.~\ref{sec:conclusion}.

\begin{figure*}[hbtp]
  \centering
  \includegraphics[width=\linewidth]{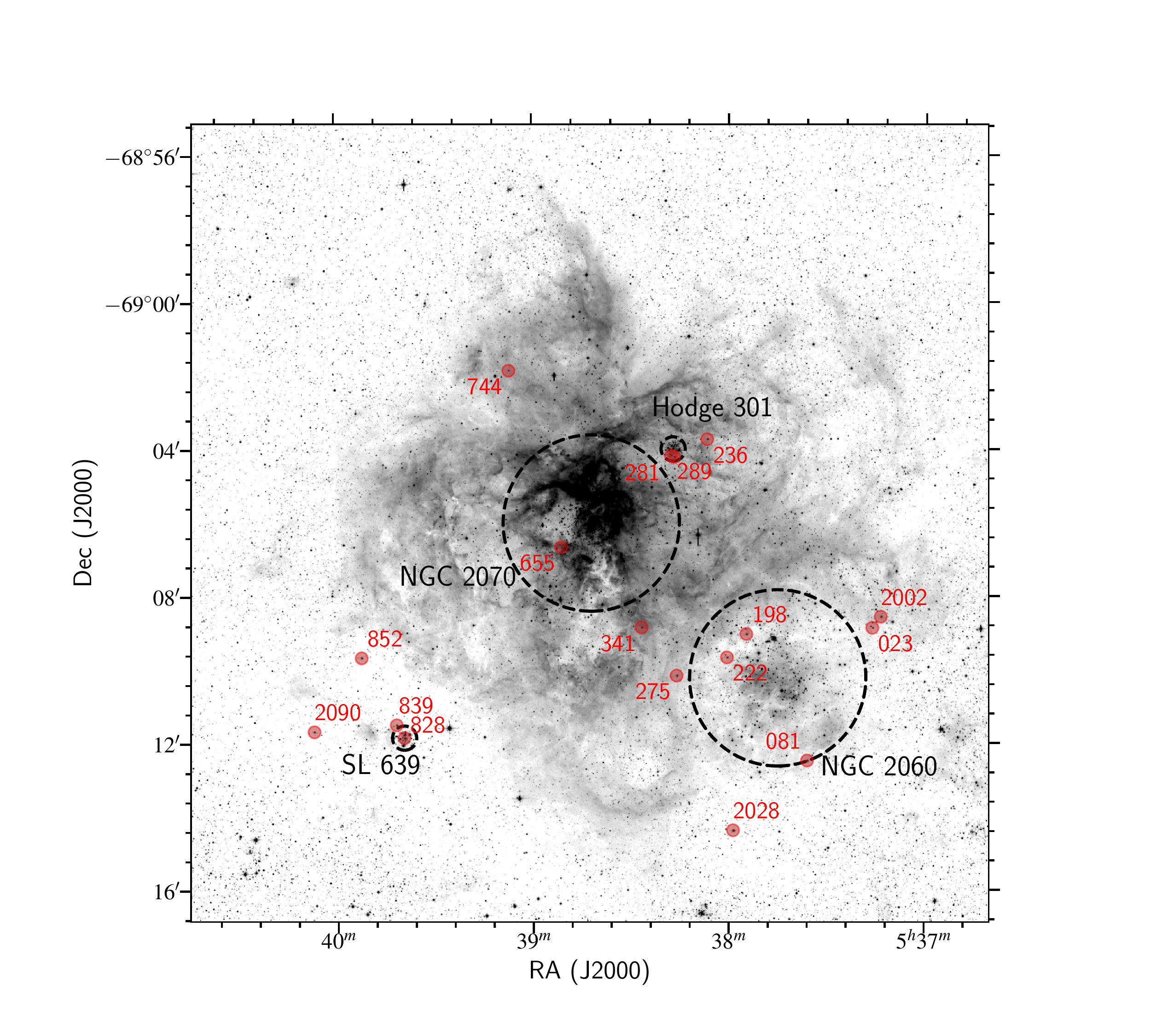}  
  \caption[]{\label{fig:30dor} %
  Spatial distribution of the cool-type stars analysed here, labelled by their VFTS identifiers. The spatial extents of NGC\,2070, NGC\,2060, SL\,639, and Hodge\,301 \citep[as adopted by][]{2015A&A...574A..13E}, are indicated by the overlaid dashed circles. The underlying image is from a $V$-band mosaic taken with ESO’s Wide Field Imager on the 2.2 m telescope at La Silla.
  Despite their location within the boundaries of the NGC\,2060, NGC\,2070 clusters, given the ages of these populations, the RSGs in these regions are considered field members.}
\end{figure*}

\section{Observations}    \label{sec:observations}

The data were obtained with the VLT-FLAMES spectrograph~\citep{2002Msngr.110....1P} as part of the Tarantula Survey. Nine multi-object fibre configurations (Fields A to I) were used to assemble the VFTS sample, 
and the wavelength coverage and resolving power for the three FLAMES-Giraffe settings used are given in Table~\ref{tb:settings}. 

The observing strategy involved two back-to-back exposures executed at the
telescope in the service queue. To ensure sufficient signal-to-noise (S/N) for quantitative spectral
analysis, three observations were obtained with the LR02 and LR03 settings and two with the HR15N setting. Three additional observations were obtained with the LR02 setting to look for RV variations, with the last (sixth) epoch obtained two observing seasons later to give a $\sim$1\,yr baseline to improve the detection of long-period systems. There were no strong time constraints on the execution of the first three LR02, nor the LR03/HR15N, observations, so there is additional temporal information for some of the observed fields. For our current purposes, observations of a given target obtained on the same night were combined to improve the final S/N of the spectra. Full details of the observations are given in Appendix~A of Paper~I.

\subsection{Red supergiant sample}
Photometric criteria ($V$\,$<$\,16\,mag, $(B-V)$\,$>$\,1\,mag) were used to select the late-type, luminous stars observed by the VFTS, resulting in 18 candidate RSGs. Analysis of their physical parameters
by B19 confirms that the large majority of these are RSGs (spanning
a wide range in luminosity), but also include what appear to be
four massive stars on the asymptotic giant branch (AGB) and one
foreground star (VFTS\,793, which is removed from the current sample). Note that this sample includes
three stars with identifications of the form {\it 2xxx} -- these were drawn from objects previously excluded from the survey catalogue as potentially foreground stars or LMC members with poor S/N spectra (see Table~A.1 in B19).

The RV analysis presented here considers the same sample of VFTS objects as B19 (excluding VFTS\,793). The locations of the 17 targets in relation to the main clusters/associations in the region are shown in Figure~\ref{fig:30dor}; these include the two older clusters, Hodge\,301~\citep{1988PASP..100.1051H} and SL\,639~\citep{1963IrAJ....6...74S}, for which B19 estimated
ages of 24$_{-3}^{+5}$\,Myr and 22$_{-5}^{+6}$\,Myr, respectively.

{\small
\begin{table}
\caption{Wavelength coverage, resolving power ($R$), and number of slices used in the radial-velocity analysis for the three observed FLAMES--Giraffe settings.}
\label{tb:settings}      
\centering                                      
\begin{tabular}{lccc}          
\hline\hline                        
Setting & Range [\AA] & $R$ [$\lambda/\delta\lambda$] & No. slices \\    
\hline                                   
LR02 & 3960 -- 4564 & \z7000 & 16\\
LR03 & 4499 -- 5071 & \z8500 & 15\\
HR15N & 6442 -- 6817 & 16000 & \z9\\
\hline                                             
\end{tabular}
\end{table}
}

\section{Radial velocity analysis} \label{sec:rv}
At the resolution of the VFTS data, the high density of absorption features in the optical spectra of RSGs enables precise RVs to be estimated to better than 1\,\kms\/ in most cases. 
The RVs for our sample were estimated for each epoch using a slicing technique that splits each spectrum into small wavelength slices.
The width and location of these slices are tailored to each setting, ranging from $\sim$30\AA\/ (LR02) to 45\AA\/ (HR15N), but kept constant for all targets (see Table~\ref{tb:settings}).
Using an iterative cross-correlation approach, these slices are compared to a spectrum calculated from a {\sc marcs} model atmosphere \citep{2008A&A...486..951G} with appropriate physical parameters for a RSG, i.e., T$_{\rm eff}=$~3900\,K, log\,$g$~=~0.5. 
The final RV estimate for each epoch is the average of the cross-correlation estimates for the slices.
The precision of our measurements is estimated using a comparison between all of the slices ($\sigma/\sqrt{N_{\rm slices}}$; where $N_{\rm slices}$ is the total number of slices for the spectrum) and, in general, is better than 1\,\kms.
Average RVs for each target from all three settings are presented in Table~\ref{tb:rv}, along with the adopted line-of-sight systemic velocities ($v_{\rm 1D}$).

\begin{table*}
{\small
\caption{Estimated mean radial velocities (RVs) and associated uncertainties from the three spectrograph settings.
Adopted line-of-sight velocities ($v_{\rm 1D}$) are a weighted average of the individual epochs in the LR02 and LR03 settings.
}              
\label{tb:rv}      
\centering                                      
\begin{tabular}{ccccccl}          
\hline\hline                        
VFTS ID & Field & \multicolumn{3}{c}{RV [\kms]} & $v_{\rm 1D}$ [\kms] & Notes\\    
\cline{3-5}
& & LR02 & LR03 & HR15N\\
\hline
  0023  & I* &    270.1\,$\pm$\,0.4     &   270.8\,$\pm$\,0.8     &   270.7\,$\pm$\,0.3     &   270.3\,$\pm$\,0.3 & AGB candidate           \\ 
  0081  & B* &    284.9\,$\pm$\,0.3     & \p285.1\,$\pm$\,0.9$^b$ & \p287.1\,$\pm$\,0.3$^b$ &   284.9\,$\pm$\,0.3 &         RV\,$=$\,287.6\,\kms\/ (GF15)                \\ 
  0198  & A* &    258.2\,$\pm$\,0.3     &   257.5\,$\pm$\,0.6     &   260.1\,$\pm$\,0.2     &   258.1\,$\pm$\,0.3 &     RV\,$=$\,260.3\,\kms\/ (GF15)                     \\ 
  0222  & C* &  \p253.2\,$\pm$\,0.3$^a$ & \p254.4\,$\pm$\,0.9$^b$ &   256.1\,$\pm$\,0.3     &   253.3\,$\pm$\,0.3 & AGB candidate           \\ 
  0236  & C* &  \p261.2\,$\pm$\,0.3$^a$ & \p261.7\,$\pm$\,0.8$^b$ &   263.6\,$\pm$\,0.3     &   261.2\,$\pm$\,0.2 &    H\,301 candidate; RV\,$=$\,261.7\,\kms\/ (GF15)                \\ 
  0275  & C* &  \p289.4\,$\pm$\,0.6$^a$ & \p290.1\,$\pm$\,1.8$^b$ &   295.8\,$\pm$\,1.6     &   289.5\,$\pm$\,0.6 &  RV\,$=$\,289.4\,\kms\/ (MO03)                       \\ 
  0281  & C* &  \p263.4\,$\pm$\,0.2$^a$ & \p263.2\,$\pm$\,1.0$^b$ &   265.6\,$\pm$\,0.4     &   263.4\,$\pm$\,0.2 &    H\,301 member          \\ 
  0289  & E\phantom{*} &    260.7\,$\pm$\,0.3     & \p261.6\,$\pm$\,1.0$^b$ & \p262.7\,$\pm$\,0.5$^b$ &   260.8\,$\pm$\,0.3 &    H\,301 member          \\ 
  0341  & D\phantom{*} &    282.0\,$\pm$\,0.3     &   282.3\,$\pm$\,0.6     &   285.0\,$\pm$\,0.3     &   282.1\,$\pm$\,0.3 &        RV\,$=$\,275.7\,\kms\/ (GF15)                  \\ 
  0655  & I* &    284.4\,$\pm$\,0.4     &   285.1\,$\pm$\,0.7     &   286.8\,$\pm$\,0.3     &   284.6\,$\pm$\,0.3 & AGB candidate           \\ 
  0744  & C* &  \p250.6\,$\pm$\,0.4$^a$ & \p253.9\,$\pm$\,1.1$^b$ &   255.3\,$\pm$\,0.9     &   250.9\,$\pm$\,0.4 & AGB candidate           \\ 
  0828  & B* &    249.3\,$\pm$\,0.4     & \p249.3\,$\pm$\,1.1$^b$ & \p251.0\,$\pm$\,0.7$^b$ &   249.3\,$\pm$\,0.4 & SL639 member            \\ 
  0839  & A* &    251.4\,$\pm$\,0.3     &   250.2\,$\pm$\,0.7     &   251.9\,$\pm$\,0.4     &   251.1\,$\pm$\,0.3 & SL639 member \\ 
  0852  & F\phantom{*} &    247.7\,$\pm$\,0.4     &   248.8\,$\pm$\,0.8     &   248.5\,$\pm$\,0.5     &   247.9\,$\pm$\,0.3 & SL639 candidate         \\ 
  2002  & I* &    287.5\,$\pm$\,0.4     &   287.1\,$\pm$\,0.8     &   289.2\,$\pm$\,0.7     &   287.5\,$\pm$\,0.3 &                         \\ 
  2028  & H\phantom{*} &    276.3\,$\pm$\,0.6     & \p275.7\,$\pm$\,1.2$^b$ & \p278.4\,$\pm$\,0.5$^b$ &   276.2\,$\pm$\,0.5 &                         \\ 
  2090  & I* &    249.4\,$\pm$\,0.4     &   250.6\,$\pm$\,0.7     &   253.6\,$\pm$\,0.4     &   249.7\,$\pm$\,0.4 & SL639 candidate         \\ 
  
\hline                                             
\end{tabular}
\tablefoot{$^{(a)}$Observations of Field~C on 2009-10-08 were discarded as a result of low S/N.
$^{(b)}$Estimates from only one epoch.\\
Published RVs are indicated in the final column from \citet[MO03,][]{MasseyOlsen03} and \citet[GF15,][]{2015A&A...578A...3G}.
*Estimates corrected for instrumental variation}
}
\end{table*}

In general we find excellent agreement between the LR02 and LR03 settings, however, there appears to be a systematic offset in the HR15N estimates (with a mean difference of 2.1\,\kms) cf. those from the LR02 and LR03 settings. The reason for this discrepancy is unclear, but is assumed not to be astrophysical in origin. Given this discrepancy, we do not include the HR15N results further
in our analysis (unless explicitly specified) and the adopted $v_{\rm 1D}$ is the weighted average of the LR02 and LR03 values. Although potentially greater precision is available from the (higher-resolution) HR15N data, in addition to the apparent offset, there were fewer observations with this setting and the S/N tends to be lower.

Comparisons of the RVs estimated for hot stars from the VFTS can provide an independent check of our results as they should trace the same stellar population. The mean velocities for the RSGs (this study) and B-type stars \citep{2015A&A...574A..13E} for the different associations in the 30~Dor region are given in Table~\ref{tb:rv_compare}. For consistency, we adopted the same definitions (central coordinates and radii) as those used by \citet{2013A&A...550A.107S} and \citet{2015A&A...574A..13E}, and we included all the candidate members of the older clusters from Table~\ref{tb:rv}. 

\begin{table*}
\caption{Mean radial velocities (RVs) and dispersions of the Hodge\,301 and SL\,639 clusters compared to the local field population for RSGs (this study) and B-type stars in the VFTS \citep{2015A&A...574A..13E}.}              
\label{tb:rv_compare}      
\centering                                      
\begin{tabular}{lcccccccc}          
\hline\hline                        
Region & $\alpha$ & $\delta$  & \multicolumn{2}{c}{B-type} & & \multicolumn{2}{c}{RSGs}\\    
\cline{4-5}  \cline{7-8} 
& \multicolumn{2}{c}{[J2000]} & No. of stars  & RV\,$\pm$\,$\sigma$ & & No. of stars  & RV\,$\pm$\,$\sigma$\\
\hline
  All  		& \ldots & \ldots & 298 & 270.4\,$\pm$\,12.4 & & 17 & 265.2\,$\pm$\,14.8 \\
\hline
  Hodge\,301& 05 38 17.0 & $-$69 04 01.0 & \phantom{0}14 & 261.8\,$\pm$\,5.5\phantom{0}  & & \phantom{0}3 &  262.1\,$\pm$\,1.4\phantom{0}\\
  SL\,639	& 05 39 39.4 & $-$69 11 52.1 & \phantom{0}11 & 253.5\,$\pm$\,3.9\phantom{0}  & & \phantom{0}4 & 249.9\,$\pm$\,1.0\phantom{0}\\
  Field		& \ldots     & \ldots 		 & 173& 271.6\,$\pm$\,11.8 & & 10 & 273.3\,$\pm$\,14.0\\
\hline                                             
\end{tabular}
\tablefoot{Central coordinates as defined by~\citet{2015A&A...574A..13E}. For these estimates all candidate members of SL\,639 were assumed as members.}
\end{table*}

\begin{figure}[hbtp]
  \centering
  \includegraphics[width=\columnwidth]{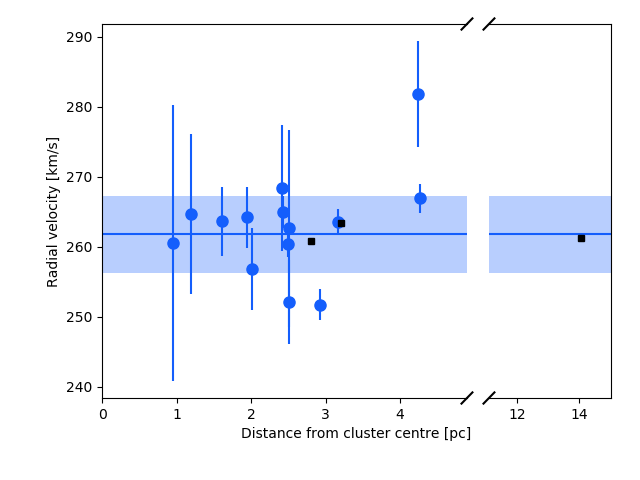}  
  \caption[]{\label{fig:H301RV} %
  Radial velocity (RV) estimates for members of Hodge\,301.
  Results for the B-type stars from \citet{2015A&A...574A..13E} are shown by the blue circles and estimates for the RSGs in this study are shown by the black squares (for which the uncertainties are not
  visible on this scale).
  The average and standard deviation of the B-type stars are indicated by the blue line and shaded region, respectively. The systemic velocity estimated from the RSGs is in excellent agreement with that from the hotter stars (see also Table~\ref{tb:rv_compare}).}
\end{figure}

\begin{figure}[hbtp]
  \centering
  \includegraphics[width=\columnwidth]{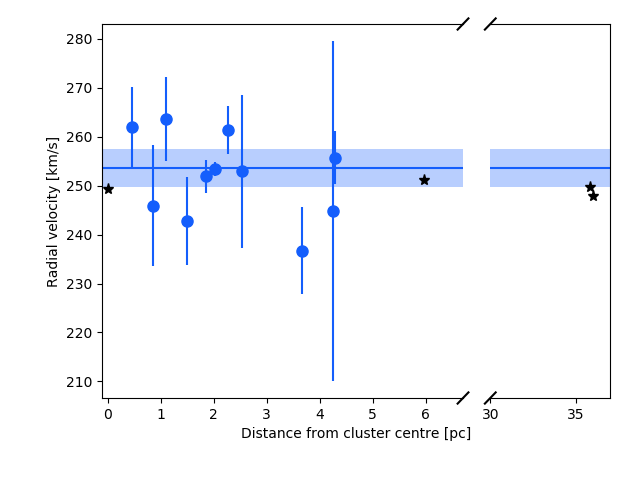}  
  \caption[]{\label{fig:SL639RV} %
  As Figure~\ref{fig:H301RV}, for the four RSGs in SL\,639 (shown with black stars).
  }
\end{figure}

The agreement between the estimates for hot and cool stars in the different groups in Table~\ref{tb:rv_compare} is very good (within 1$\sigma$ in all cases), providing independent support of our adopted RV method for the cool stars.
We note that here and in Table~\ref{tb:rv}, we used unbiased estimates of the standard deviation that account for the small sample sizes.
Figures~\ref{fig:H301RV} and~\ref{fig:SL639RV} show the RVs for the individual B-type stars \citep[from][]{2015A&A...574A..13E} compared with our estimates for the RSGs in Hodge\,301 and SL\,639, respectively. Within the uncertainties the systemic velocities are in good agreement. Moreover, the dispersion of the RSGs appears smaller. We expect this is a consequence of the large number of spectral lines available in the RSG spectra (cf. the more limited number of helium and metallic lines in B-type spectra), combined with lower rotational broadening in the cool stars (which limits the RV precision in the more rapidly rotating B-type stars).

Five of our targets have published estimates in the literature, as indicated in the final column of Table~\ref{tb:rv}. Four of these are from the study of \citet{2015A&A...578A...3G}, with a quoted uncertainty of $\pm$4\,\kms\/ on their estimates. Within the respective uncertainties these are also in good agreement with our results.

We find no evidence of runaway RSGs, with radial velocities offset by more than 30\,\kms\ 
compared to the surrounding population \citep{1961BAN....15..265B,2011MNRAS.414.3501E,2018MNRAS.477.5261B,2019MNRAS.482L.102R}.
Based on the binary supernova scenario, \cite{2019MNRAS.482L.102R}, find that the the number of runaway stars is dwarfed by an order of magnitude by the number of slower moving -- nevertheless unbound -- walkaway stars.
Through dynamical simulations of young ($<$3\,Myr) star clusters~\citep{2016A&A...590A.107O} suggest that a significant fraction of their stellar population can be ejected through dynamical interactions, resulting in a similar velocity distribution for ejected stars.

This walkaway scenario is a possible explanation for the location of the RSGs near Hodge\,301 and SL\,639.
Particularly in the case of SL\,639, as the stars appear to be genuine cluster members (see Section~\ref{sec:Mdyn}), it is unlikely that these stars formed in their current location.

\subsection{Multi-epoch analysis} 
\label{sub:multi}

\begin{figure*}[hbtp]
  \centering
  \includegraphics[width=\linewidth]{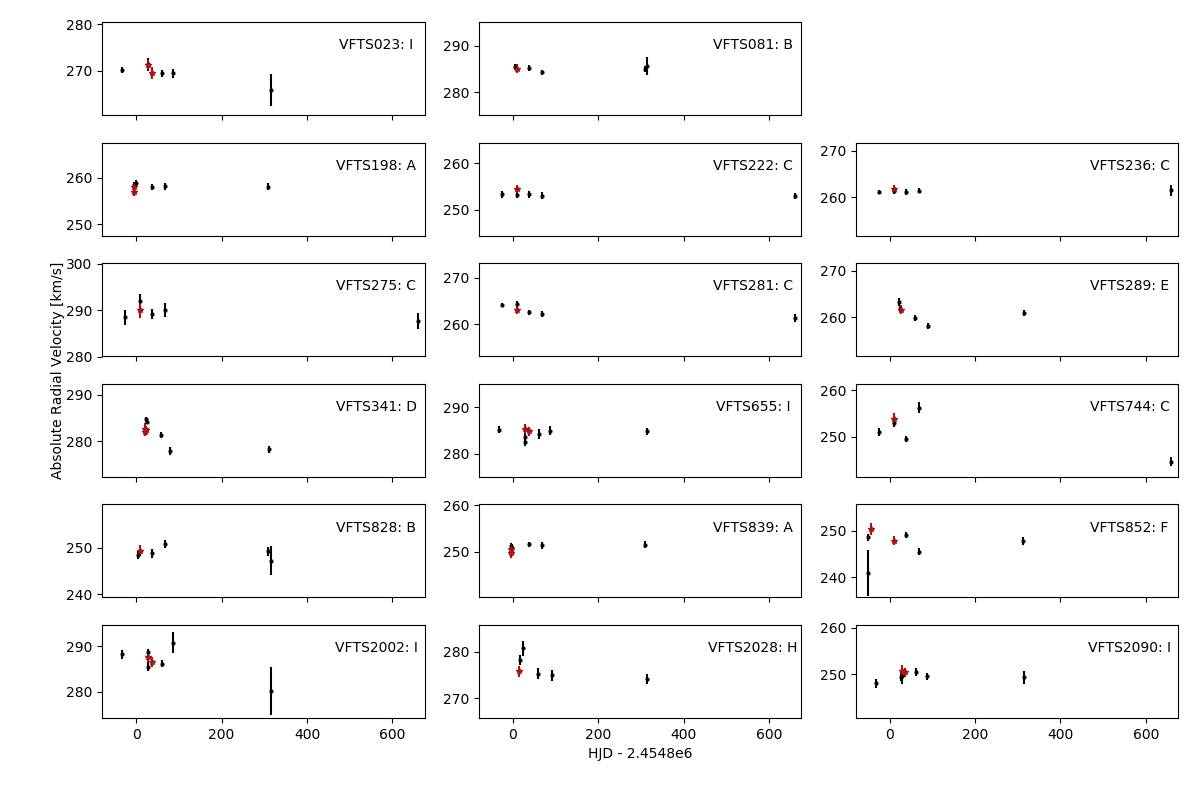}  
  \caption[]{\label{fig:rvall} %
  Absolute RV estimates from the multi-epoch observations of our sample as a function of Heliocentric Julian Date (HJD).
  Black points are the LR02 measurements and red stars show the LR03 measurements, note the excellent agreement between the two settings (see Table~\ref{tb:rv} for details). 
  Each panel has the same vertical scale to highlight the amplitude of the observed variations. The fibre configuration (Fields A-I) in which each target was observed is also indicated. Corrections for instrumental variation have been implemented for 13 of the targets (excl. VFTS\,289, 341, 852, 2028; see text for details).}
\end{figure*}

\begin{figure}[hbtp]
  \centering
  \includegraphics[width=\columnwidth]{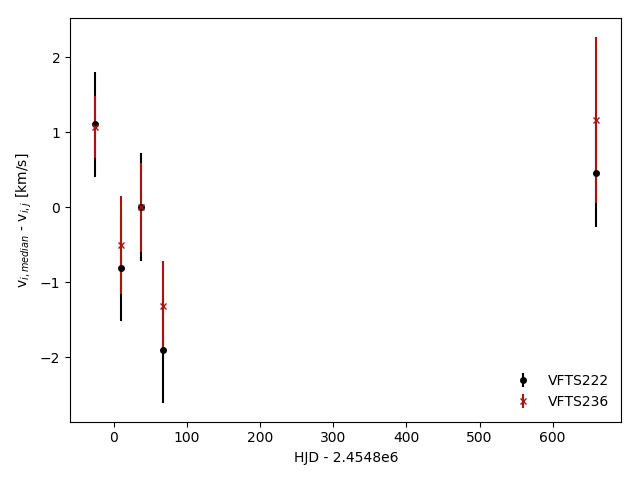}  
  \caption[]{\label{fig:222vs236} %
  Relative radial velocities from the LR02 observations of two targets in the Field~C fibre configuration: VFTS\,222 (black dots) and VFTS\,236 (red crosses). 
  The median value for each star has been subtracted to highlight the similarities between the epochs. This suggests an instrumental variation in the spectrograph. 
  The two targets shown here define the correction for Field C.}
\end{figure}

Initial tests showed a correlation between the velocities estimated for targets observed in the same fibre configuration. Table~\ref{tb:rvall} lists the complete set of RV estimates for our sample, and Figure~\ref{fig:rvall} shows the RV curves for each target. Our targets were observed within eight of the nine fibre configurations used for the VFTS, with some observed in the same field (see Table~\ref{tb:rv} for details). From initial analysis of the RV estimates we noticed that targets observed in the same fibre configuration had very similar relative velocities. An example of this correlation is shown in Figure~\ref{fig:222vs236} for VFTS\,222 and 236, both observed in Field~C, and also highlights the scale of the instrumental variation (on the order of $\pm$2\,\kms.)

There is no plausible explanation of why the RVs of different RSGs should be coupled in such
a way, and we concluded that this is linked to the wavelength calibration of the data. The measured variations correlate approximately with differences in the on-instrument temperature at the time of the arc calibrations (taken during the day) and the science observations; unfortunately no consistent relationship could be found when considering
the recorded temperatures. Where possible we have therefore attempted to calibrate the velocity scale empirically, via groups of RV measurements within the different VFTS fields. In doing this we acknowledge that these stars may also show physical variations at a level comparable to the (presumed) instrumental effect, so we potentially remove some genuine RV variations for a given target.

To identify such instrumental variations, the RVs of each target (cf. their medians) are compared with those for other targets in the same fibre configuration.\footnote{An attempt was made to use B-type stars observed in the same configuration to estimate this offset, however, the precision on the RV measurements for these stars is not sufficient to provide a conclusive result.}
We find significant variability (using Pearson's correlation coefficient) for at least one pair of targets
for each field. The correction applied to the results for each epoch is then the weighted average of the relative velocities for a pair of targets (or three for Field~I) within that field. These corrections were then implemented to each target of the relevant field, resulting in corrected RVs for 13 stars. No correction was possible for the remainder (VFTS\,289, 341, 852, and 2028) as only one star from our sample was observed per fibre configuration.

After correction, all of the targets are consistent with an absence of significant RV variation on scales above 5\,\kms, except for VFTS\,744, which displays a variation of 11.6\,\kms\/ between epochs, 
comparable to the variation found by~\citet{2007A&A...469..671J} for several of their targets (see Section~\ref{sub:phot_var} for more discussion of the variability of VFTS\,744).
The remaining targets that display differences on this scale between epochs all have a combination of larger than average uncertainties or/and uncorrected instrumental variation (e.g. VFTS\,023, 852 and 2002, see Figure~\ref{fig:rvall}).
No apparent periodicities were detected in the spectroscopy (as found for $\mu$ Cep by \citeauthor{2007A&A...469..671J}), although robust checks would require more intensive spectroscopic monitoring.

It is important to discriminate, if possible, between atmospheric variations arising from large-scale convective cells that dominate the surface of RSGs~\citep{2009A&A...506.1351C, 2010A&A...515A..12C} and the effect of binary motions. 
Any RV variability from binarity would be somewhat hidden and perturbed by the rising/sinking motion of convective cells on the surface of the star that causes intrinsic velocity variability in groups of lines formed in the same part of the atmosphere~\citep[of which the variation can be up to $\sim$25\,\kms;][however the effect on the measured RV of the star effectively averages out and will be significantly less]{2007A&A...469..671J}.
We note that the slice technique used to estimate the RVs is, by its nature, particularly insensitive to variations dominated by particular spectral features (or groups of features).

The low-level RV variations observed in our sample are broadly consistent with what we expect from the atmospheres of single or very long-period RSGs dominated by several large-scale convective cells~\citep[as initially hypothesised by][and shown through radiative hydrodynamical simulations by~\citealt{2002AN....323..213F}]{1975ApJ...195..137S}.
Large variations in the RVs of our sample would therefore be characteristic of significant binary motion (see Section~\ref{sec:bin_anal}). From the few known RSGs within binary systems, RV variations of order 40\,\kms\/ are seen over a full period~\citep[e.g. K1\,$=$\,20\,\kms;][]{1977JRASC..71..152W,2007AJ....133.2669E}. With orbital periods up to $\sim$20\,yr~\citep{1966ApJ...144..672P,1977JRASC..71..152W}, 
this implies RV variations of up to 6\,\kms/yr.

\subsection{Photometric variability} 
\label{sub:phot_var}

Four of our sample stars (VFTS\,023, 222, 655, 744) have been observed as part of the OGLE survey~\citep{2009AcA....59..239S}.
The level of photometric variability for all these targets is in good agreement with the expected level of variation for this type of long period semi-regular variable stars~\citep{1983ApJ...272...99W}.

This type of variability is thought to arise from radial pulsations~\citep[e.g.][]{1997A&A...327..224H,2010ApJ...725.1170S} and is known to exist in various Galactic RSGs at this level and on periods of several hundred days~\citep{2006MNRAS.372.1721K}, as well as longer-term periodic variations~\citep{2006MNRAS.372.1721K,2008AJ....135.1450G}.

For all of the four targets with OGLE data available, the Lomb-Scargle Periodogram was searched for potential periodicities.
Only for VFTS\,744 was there any suggestion of significant periods, this supports the argument that VFTS\,744 is an AGB star (see B19 and Section~\ref{sec:bin_anal} for details).

\section{Physical properties of the clusters}
\label{sec:Mdyn}

Given the good agreement of our RV estimates with previous values and the apparently small velocity dispersions in Hodge\,301 
and SL\,639, we investigated the physical properties of the two clusters
following the approach used for NGC\,2100, a RSG cluster close to 30~Dor, by \citet{2016MNRAS.458.3968P}.

We estimated the mean cluster velocity and line-of-sight velocity dispersion using an implementation of the Markov chain Monte Carlo~\citep[MCMC; ][]{2010CAMCS.5..65G} method, where {\sc emcee}~\citep{2013PASP..125..306F} is used to sample from the posterior probability distribution.
The likelihood function used here is the same used by~\citet{2016MNRAS.458.3968P}.

Using the line-of-sight velocity dispersion of the cluster, the Virial theorem can be used to define the dynamical mass of the cluster $M_{dyn}$ as,

\begin{equation}
  M_{dyn} = \frac{\eta\sigma_{1D}^{2}r_{\rm eff}}{G}
  \label{eq:vir}
\end{equation}

\noindent where $\eta$~=~9.75 for a bound cluster and $r_{\rm eff}$ is the effective radius of the cluster.

To calculate the cluster properties we include all RSG candidates for each cluster as well as the non-variable B-type stars from~\citet{2015A&A...574A..13E}.
All B-type stars flagged as variable are excluded as binary motion has the effect of artificially increasing the cluster velocity dispersion~\citep{2010MNRAS.402.1750G}.

Some justification must be given for the inclusion of all RSG candidate cluster members, particularly as some of these candidates are located up to $\sim$35\,pc from the cluster centre. 
We can see from Table~\ref{tb:rv_compare} that the dispersion of the field population of both the RSGs and B-type stars is $\sim$13\,\kms\/ and the dispersion of the clusters is (from RSG results alone) an order of magnitude lower (see Table~\ref{tb:sig_compare}). 

To assess the status of the outlying stars, we calculated
the probability of their measured velocities being consistent with membership of the clusters cf. membership of the local field population. The parent
population was defined by a convolution of two Gaussian functions, where
$P(x|\{\mu, \sigma\}_{Cluster}) + P(x|\{\mu, \sigma\}_{Field}) = 1$, and with
the systemic velocities and dispersions of Hodge\,301, SL\,693 and the local field taken from~\citet{2015A&A...574A..13E}.
In all cases, we find a probability greater than $\sim$90\% that the targets are cluster members rather than part of the local field distribution.
Therefore, despite their distance from the cluster centres, based on their kinematic properties we classify VFTS\,236 as a member of Hodge\,301, and VFTS\,852 and 2090 as members of SL\,639.

Given the small sample size of the RSGs in each cluster, while our estimates take into account the sample size, the velocity dispersions and mass estimates from solely RSGs may be better treated as upper limits. Combining the RSG results with those of the B-type stars, allows us to provide more stringent constraints on these measurements.

\subsection{Hodge\,301} \label{sub:h301}

From the three RSGs and 14 B-type members, we estimate the mean cluster velocity of Hodge\,301 as 262.1\,$\pm$\,1.4\,\kms, with a line-of-sight dispersion of 4.2\,$^{+1.7}_{-1.2}$\,\kms.

\citet{2016ApJ...833..154C} found that a 4\,pc radius included 85\% of the stars within the cluster. Assuming a Plummer density law~\citep{1911MNRAS..71..460P}, where one solves for the mass contained within a radius $M(r)$, we find that the radius containing 50\% of the stars (i.e. the effective radius), is $r_{\rm eff}$~=~1.68\,pc.
This assumes that the half-light radius is equivalent to the half-mass radius ~\citep[although this is not the case for mass-segregated clusters, see][]{2008MNRAS.391..190G,2010MNRAS.408L..76F}.

Our estimated dynamical mass of Hodge\,301, from the combination of RSGs and B-type stars, is $\log (M_{\rm dyn}/M_{\odot})=$~3.8\,$\pm$\,0.3.
This is in excellent agreement with the photometric estimate of $\log (M_{\rm phot}$/M$_{\odot}$) = 3.9 from \citet{2016ApJ...833..154C}.

\subsection{SL 639} \label{sub:sl639}
Using the same methods for SL\,639 (from the four RSGs and 11-B-type stars), we estimated a mean cluster velocity of 249.9\,$\pm$\,1.0\,\kms, with a line-of-sight dispersion of 1.9\,$^{+1.8}_{-1.1}$\,\kms.
In absence of published structural parameters, we assume $r_{\rm eff}$\,$=$\,1.68\,pc as per Hodge\,301. From inspection of the available imaging SL\,639 appears less extended than Hodge\,301, so adopting this effective radius in eqn.~\ref{eq:vir} gives an upper limit on the dynamical mass of SL\,639 of
$\log (M_{dyn}/M_{\odot}$)~<~3.1\,$\pm$\,0.8. This is the first mass estimate for this cluster.

\begin{table}
\caption{Estimated line-of-sight velocity dispersion and dynamical masses from RSGs and B-type stars in the Hodge\,301 and SL\,639 clusters.}              
\label{tb:sig_compare}      
\centering                                      

\begin{tabular}{l cc}          
\hline\hline      
 & $\sigma_{\rm 1D}$ [\kms]  & $\log (M_{\rm dyn}/M_{\odot})$\\

\hline
Hodge\,301 \\
~~~~RSG    & 2.1\,$^{+3.7}_{-1.4}$ & \phantom{<}3.2\,$\pm$\,1.0\\
~~~~B-type & 5.8\,$^{+2.5}_{-1.7}$ & \phantom{<}4.1\,$\pm$\,0.3\\
~~~~All    & 4.2\,$^{+1.7}_{-1.2}$ & \phantom{<}3.8\,$\pm$\,0.3\\
\hline
SL\,639 \\
~~~~RSG   & 1.4\,$^{+1.9}_{-0.9}$ & $<$2.9\,$\pm$\,0.9\\
~~~~B-type & 3.1\,$^{+3.7}_{-2.3}$ & $<$3.6\,$\pm$\,1.0\\
~~~~All    & 1.9\,$^{+1.8}_{-1.1}$ & $<$3.1\,$\pm$\,0.8\\
\hline                                             
\end{tabular}
\tablefoot{Dynamical masses for SL\,639 are upper limits as they adopt the
same effective radius as Hodge\,301 ($r_{\rm eff}$\,$=$\,1.68\,pc).
}
\end{table}

\subsection{On the velocity dispersion of young massive clusters} \label{sub:sigma_1d}

As a result of the improved precision in the RSG RV measurements, one might expect that the estimates of the line-of-sight velocity dispersion from solely RSGs would provide more stringent constraints.
However, the limiting factor in both cases studied here is the sample size, which is reflected in the larger uncertainties of the RSG estimates (see Table~\ref{tb:sig_compare}).
The limiting factor in the case of the B-type sample is that of the precision of the RV measurements and potential undetected binarity, therefore, even though the sample size is larger by a factor of three to five, the uncertainties are not significantly smaller than the solely RSG estimates.

The effects of contamination within our RSG sample from undetected binarity, pulsations or including a non-cluster member (unlikely based on the assessment of the measured RVs compared against the distribution of velocities in the field stars), would all act to increase the measured velocity dispersion.
Therefore we argue that, despite the large uncertainties in the RSG measurements, the RSG sample is not heavily contaminated by the aforementioned effects.

The striking agreement of the average velocities among the members of each cluster is further evidence for an absence of RV variability for the RSGs, otherwise the velocity
dispersions would be dominated by such variations \citep[e.g.][]{2010MNRAS.402.1750G,2011IAUS..272..474S,2012A&A...546A..73H}.
This is also confirmed by results from the LR02 spectra, where the maximum RV variation between any given epoch for all cluster members is 5.2\,\kms\/ (VFTS\,289), which is probably linked to the instrument variations discussed in Section~\ref{sec:rv}.

The results of the previous sections highlight that RSGs are effective tracers of the kinematic properties of young clusters, potentially offering advantages in RV precision over their main-sequence counterparts.
 

\section{Binary analysis} \label{sec:bin_anal}

There is growing evidence that most massive stars evolve within binary systems~\citep[e.g.][]{2012Sci...337..444S,2013A&A...550A.107S,2013ARA&A..51..269D,2014ApJS..215...15S,2014ApJS..213...34K,2015A&A...580A..93D,2017ApJS..230...15M}.
Simulations of how these stars evolve after the main sequence suggest that RSG lifetimes are significantly shortened as a result of Roche-lobe overflow and common-envelope evolution~\citep{2008MNRAS.384.1109E}.
In the literature there are only a handful of examples of binary systems containing RSGs, and all with orbital periods greater than several hundred days to several decades. For example, VV~Cep has an orbital period of about 20~yr and two almost equal mass companions \citep{1977JRASC..71..152W} while  QS~Vul
is the shortest-period system known with P~$=$~249\,d \citep{1993A&A...274..225G}, although
 its status as a RSG was questioned by \citet{2007AJ....133.2669E} from estimates of its stellar parameters. 

Before searching for RSG binaries in our dataset, we used Monte Carlo simulations to check  the sensitivity of our observations to long-period binaries. To do so, we computed the detection probability of RSG binaries such as VV Cep ($P=20.3$~yr, $q=1.0$, $e=0.35$), should they have been observed with a temporal sampling and RV measurement accuracy corresponding to each of the objects in our sample. We found that VV~Cep-like binaries would have been detected in 10 to 60\% of the cases -- with an average of 33\% -- depending on the data set considered, i.e. we would expect to detect at the 3-sigma variability level at least 5 binaries if all our objects where VV Cep-analogous. This illustrates that our present observations are able to detect some binary systems if they would be present among our RSG(-candidate) sample despite the limited baseline of the VFTS observations.

\subsection{RV-variability detection}

The multiplicity of the hot stars from the VFTS survey have been investigated 
in two papers. \citet{2013A&A...550A.107S} studied the multiplicity properties of the 360 O-type stars, finding an intrinsic binary fraction (after accounting for observational biases) of 51\,$\pm$\,4\,\%. In good agreement, \citet{2015A&A...580A..93D} performed a similar analysis of the early B-type stars, finding an intrinsic binary fraction of 58\,$\pm$\,11\,\%,
although they also noted that the fraction might be significantly lower in Hodge\,301 and
SL\,639. The RV variability search criteria used by these two papers to detect binarity were:

\begin{equation}
\label{eq:var}
\frac{|v_i - v_j|}{\sqrt[]{\sigma_i^2 + \sigma_j^2}} > \sigma_{lim} ~{\rm and}~ |v_i - v_j| > \Delta RV_{min};
\end{equation}

where, $v_i$, $\sigma_i$ ($v_j$, $\sigma_j$) are the measured radial velocities and corresponding uncertainty, at epoch $i$ ($j$), respectively.
$\sigma_{lim}$ represents a limit set by the uncertainties on the RV measurements on significant variation.
Both criteria need to be met by at least one pair of RV measurements ($v_i, v_j$) for the object to be considered RV variable and, hence, a spectroscopic binary candidate.
The choice of $\Delta RV_{min}$ for these studies reflected the distribution of $\Delta RV$ values estimated, 20 and 16\,\kms\/ for the O- and B-type stars, respectively.
These limits were chosen to reject false positives arising from pulsational RV variations and the choice of $\sigma_{lim}$~=~4.0 was selected to ensure no false positives given their sample sizes.
For the current study, a $\sigma_{lim}$~=~3.0 is selected, which reflects the smaller sample size and excludes the presence of any false positives at the 99.7\% level.

\begin{figure}[btp]
  \centering
  \includegraphics[width=\columnwidth]{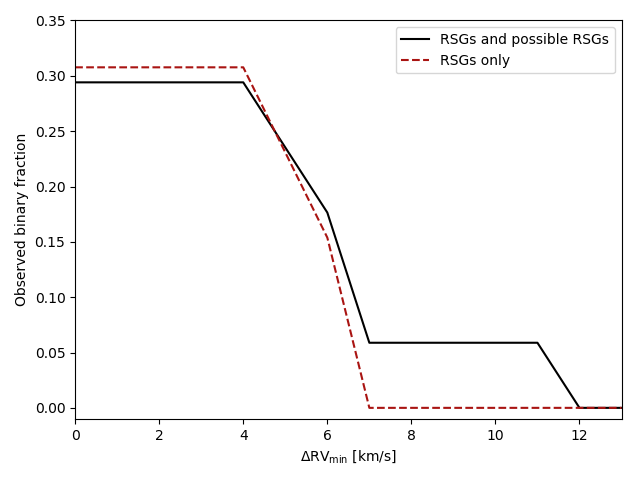}  
  \caption[]{\label{fig:bf} %
  Fraction of systems that satisfy the 3$\sigma$ variability criterion (equation~\ref{eq:var}), as a function of the $\Delta RV_{min}$ variability criterion for the entire sample (black solid line) and for probable RSGs (red dashed line; as defined in B19). 
  This figure highlights that the majority of the sample do not satisfy the 3$\sigma$ variability criterion and that the maximum observed binary fraction is $\sim$0.30.
  Only one target in the sample (VFTS\,744) displays significant variability above $\Delta$RV$_{\rm min}$~=~7.0\,\kms.
  }
\end{figure}

The choice of $\Delta RV_{min}$ is an important consideration for the present study. 
Figure~\ref{fig:bf} shows the observed binary fraction (i.e. the fraction of stars meeting both variability criteria) of the whole sample (black solid line) and probable RSGs (red dashed line) as a function of the $\Delta RV_{min}$ parameter.

To evaluate the best choice of $\Delta RV_{min}$, we investigated the typical velocities of known RSG binaries (with the caveat that the known systems are likely a biased sample of the population of RSGs in binary systems).
As mentioned above, literature examples of RSGs binaries~\citep[e.g.][]{1977JRASC..71..152W} typically contain B-type star companions~\citep{2018AJ....156..225N}, 
where the peak-to-peak RV variations of the primary are $\sim$40\,\kms~\citep[e.g. K$\sim$20\,\kms\/ for VV Cep and K$\sim$27\,\kms\/ for QS Vul;][]{1966ApJ...144..672P,1992A&AS...95..589H,2007AJ....133.2669E},
over a period of up to 20\,yr~\citep{1977JRASC..71..152W}, or up to $\sim$6\,\kms\/ per year.

It is informative to consider the semi-amplitude velocity ($K$) of the RSG that is expected for a range of periods and mass ratios, taking into account the minimum periods excluded by the size of the star. In Appendix~\ref{ap:kcalc} we discuss the cases of an 8 and 15\,M$_{\odot}$ RSG and, adopting the latter as typical of our sample, 
we find that in the range $q >$~0.1, 3.25~$< \log$P [day]~$<$~4.25, we expect 2\,$<$\,K\,$<$\,30\,\kms.
Clearly systems with $K$$\sim$2\,\kms\ would go undetected  for $\Delta RV_{min}$ greater than this value (cf Figure~\ref{fig:bf}),
therefore we adopt a more conservative $\Delta RV_{min}$ limit for our data of $<$4.0\,\kms.

Adopting $\Delta RV_{min} <$~4.0\,\kms\ results in the maximum observed binary fraction for our sample. In this case, the value of $\sigma_{lim}$ for the first variability criterion is the limiting factor, and adopting $\sigma_{lim}$\,$=$\,3 we obtain a binary fraction of 0.3.
This estimate is effectively an upper limit on the observed binary fraction of our sample as our results may include false positives (from intrinsic RV variations arising in the atmospheres of RSGs or RV shifts from uncorrected instrumental variations). It is interesting to note that even when the variability criteria reach the most relaxed values, not all of the targets meet the criteria, despite the expectation that single RSGs can vary with amplitudes up to $\sim$5\,\kms\ \citep{1975ApJ...195..137S,2002AN....323..213F,2007A&A...469..671J,2010ApJ...725.1170S}. This reflects the finite precision of our RV measurements.

The sub-samples used in Figure~\ref{fig:bf} represent a luminosity cut in the sample, where the highest luminosity targets are defined as the most probable RSGs (red dashed line in Figure~\ref{fig:bf}).
Table~\ref{tb:rv} highlights targets that are considered AGB candidates by B19. Excluding these uncertain RSGs, and adopting the same RV criteria as above, we still recover an upper limit to the binary fraction of 0.3.

\subsection{Orbital configurations}
To quantitatively assess the parameter space where our data is sensitive as well as the possible orbital properties of undetected binary systems,
we investigate which orbital configurations (period, companion mass) would yield RV signals  compatible with our observational constraints using Monte Carlo simulations tuned for each target.
The parameter space considered in these simulations is orbital period of
2.7~$< \log P$[day]$< $~5.0, in steps of 0.1 (roughly corresponding to 1.5 to 275\,yr) and secondary masses are in the range $-$0.3~$< \log M_{\odot} <$~1.5, in steps of 0.1 (roughly corresponding to 0.5 to 35\,M$_{\odot}$).
Masses for the targets are taken from B19 and a flat eccentricity distribution is assumed in the range 0-0.95.
Using circular obits increases the detection probability at shorter period, but decreases it at period larger than about $10^4$ days. The impact is however limited to a few per cent such that the overall picture is left unaffected.
To optimize our sensitivity in these simulations we included all RV measurements, including those from the HR15N setting (accounting for the offset of 2.1\,\kms, see Section~\ref{sec:rv}). 

For the targets that display RV variability, these simulations allow us to identify orbital configurations that would fulfill the adopted variability criteria. For targets that do not meet the variability criteria, they allow us to place limits on companions and period ranges that should have been detected, and thus assess the properties of systems that would remain undetected given the current data.

\begin{figure}[btp]
  \centering
  \includegraphics[width=\columnwidth]{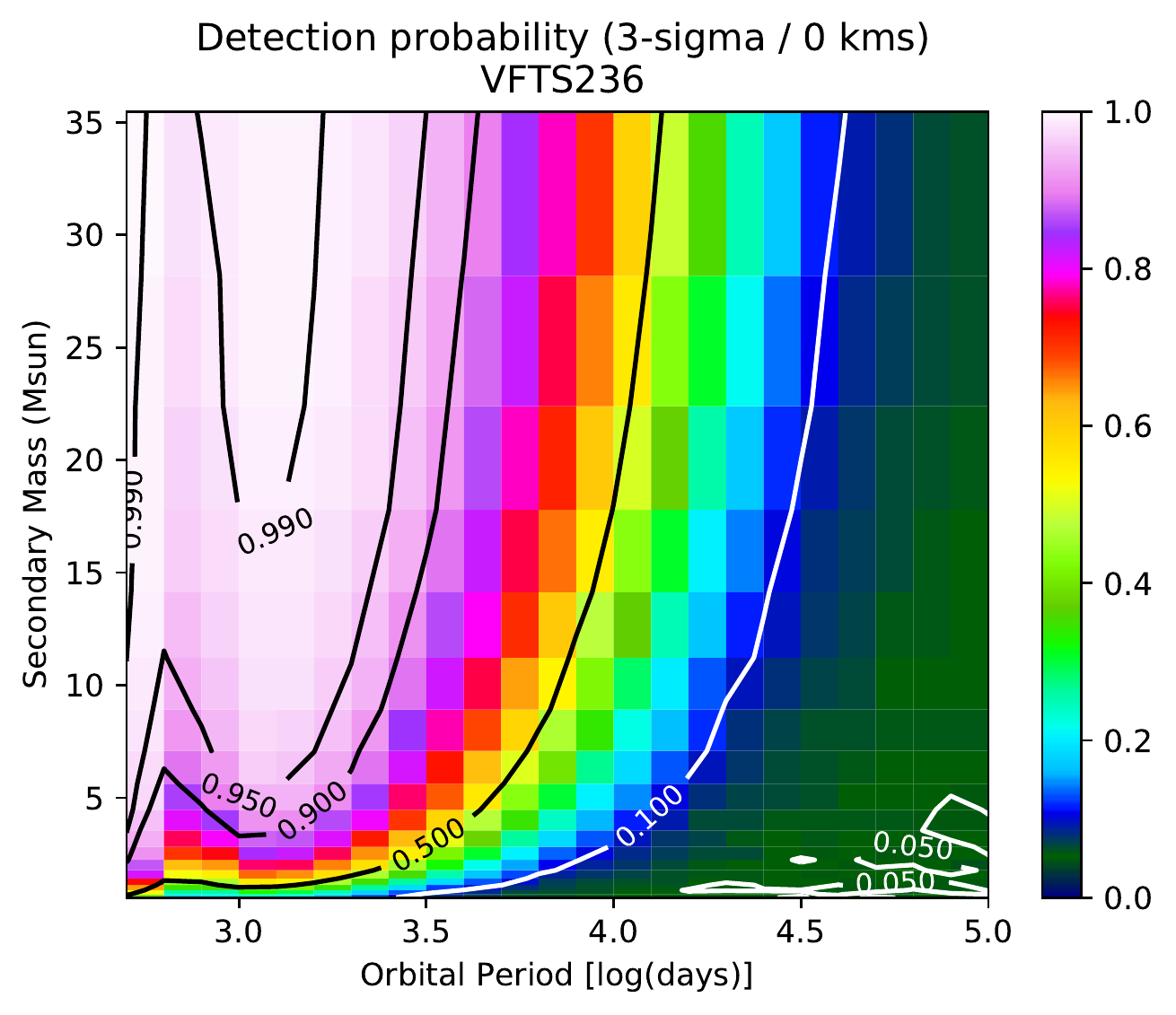}  
  \caption[]{\label{fig:236sim} %
  Population of simulated binary systems that fulfill the detection criteria of equation~\ref{eq:var}, given the observations of VFTS\,236 for $\sigma_{\rm lim}$~=~3.0\,\kms, $\Delta$RV$_{\rm min}$~=~0.0\,\kms\/ (i.e. the weakest constraints considered in Figure~\ref{fig:bf}).
  Each grid point displays the result of 10\,000 simulations where the scale is determined by the fraction of systems that meet the variability detection criteria.
  Contours indicate the 1, 5, 10, 50, 90, 95, 99 detection percentiles. 
  The mass of VFTS\,236 is 15.5\,M$_\odot$ (B19), therefore, we can reject systems with orbital periods $\log$~P[days]~$<$3.2, based on the size of the RSG primary.
  From the simulations we can reject orbital periods $\log$~P[days]~$<$3.5 for mass ratios approaching unity at the 90\% confidence level.
  Binary systems with longer orbital periods or lower mass ratios would be able to reproduce the observations.
  }
\end{figure}

Figure~\ref{fig:236sim} illustrates the results of the simulations for VFTS\,236, one of the targets that does not meet the least demanding variability criteria ($\sigma_{lim}$~=~3.0, $\Delta$RV$_{min}$~=~0.0\,\kms).
For each grid point, 10\,000 simulations of the RV variability from the resulting binary systems are evaluated using the variability criteria ($\sigma_{lim}$~=~3.0, $\Delta$RV$_{min}$~=~0.0\,\kms).
This figure is colour coded according to the number of variability detections for the 10\,000 simulations for each grid point, in addition the contours mark the 1, 5, 10, 50, 90, 95 and 99 detection percentiles.

The results of these simulations show that the strongest constraints can be placed on targets in field C, as a result of the longer baseline.
For targets that do not meet the variability criteria (i.e. apparently single stars),
we can typically exclude binary systems within the ranges q~$>$~0.3, $\log$\,P\, [days]~$<$~3.5, as these should have been detected in 90\%\ of cases (see Figure~\ref{fig:236sim}).
The period limit derived is slightly more stringent than the limit of $\log$\,P\, [days]~$\approx 3.3$  (for all mass ratios) that result from the size of the RSG (see Appendix~\ref{ap:kcalc}).

\subsection{Binary fraction}
Of the targets that meet the variability criteria ($\sigma_{\rm lim}$~=~3.0\,\kms, $\Delta$RV$_{\rm min}$~=~0.0\,\kms), the variability in VFTS\,289, 341, 852 and 2028 can likely be explained through uncorrected instrumental variation.
In fact, these are the only systems that have no correction for instrumental variations.
The only target that meets the criteria where the variation can not be explained through the lack of instrumental correction is VFTS\,744.
Our simulations yield  90\% confidence intervals of the probable orbital parameters of $q>0.1$ and $3.1 < \log\,P\,\mathrm{[days]} < 4.3$.

If we consider only probable RSGs that have a correction for the instrumental variation, we find a binary fraction of 0.0 using the variability criteria $\sigma_{\rm lim}$~=~3.0\,\kms, $\Delta$RV$_{\rm min}$~=~0.0\,\kms, for binary systems with parameters q~$>$~0.3, 3.2~$< \log$\,P\,[days]~$<$~3.5, at the 90\% confidence level.
The process of correcting for instrumental variation (described in Section~\ref{sec:rv}), clearly introduces biases and potential correlations into our analysis.
These effects act to decrease the number of binary system detections by correcting real velocity variation.
However, the conclusion that VFTS\,744 is the only target in the sample that displays significant variability that cannot be accounted for, is not affected.
Field C has the largest number of targets and the strongest correlation between the stars used to correct for the instrumental bias, if we restrict our binary fraction to these data, we find a binary fraction of 0.2, (with the sensitivity as quoted above), in agreement with our estimated upper limit.

As we have quantitatively assessed the parameter space where our observations are able to detect binarity, we now estimate the intrinsic binary fraction by comparing our observed data to the empirically defined binary stellar population of~\cite{2017ApJS..230...15M}.
Namely a flat orbital period distribution between 3.3\,$<$\,$\log$\,P [days]\,$<$\,4.3, where 50\% of companions have q\,$>$\,0.3 and the remaining in the range 0.1\,$<$\,q\,$<$\,0.3.
The mass ratio exponent is defined based on q, $-$1.7 for q\,$>$\,0.3 and $-$0.2 for 0.1\,$<$\,q\,$<$\,0.3.
Eccentricity is defined with an exponent of +0.8 within the range 0~$<$~\textit{e}\,$<$\,0.8.
With these distributions, we simulate the observed data with a binary fractions (\textit{bf}) in the range 0.0 $<$~\textit{bf}~$<$~1.0 in steps of 0.1.

We limit ourselves to the RSG sample and compare the simulations with the results of Figure~\ref{fig:bf}. Although the dispersion is large, we find that an intrinsic binary fraction of 0.3 in the period range 3.3\,$<$\,$\log$\,P [days]\,$<$\,4.3 best reproduces the sharp drops in the variability rate at $\Delta$RV$_{\rm min}$\,$\approx$\,7.0\,\kms.
This value is in agreement with the detection of long period binary system in O-type stars~\citep{2014ApJS..215...15S} and the derived binary fraction of 30\% per order of magnitude for such long period systems as derived by~\citet{2017ApJS..230...15M}.
This statistic is chosen to compare the observations with the models as below this value, the observed velocities are likely affected significantly by intrinsic variability and measurement errors.

As well as the RV variation of individual RSGs, the dispersion of RSG RVs within the cluster environments can in principle be used as an tracer of binarity~\citep[e.g.][]{2017A&A...599L...9S}. 
Simulations of the distribution of binary systems that could reproduce the observed data within the cluster proved to have little constraining power on the underlying stellar population, likely as a result of small number statistics (in both the number of targets and epochs).
More constraining power is potentially available if one considers the relative velocities of these targets as a whole.
This is beyond the scope of the current sample, but will be partly addressed by a study of NGC\,330
in the SMC (Patrick et al. in prep.). 

The lack of any significant RV variables in the current study lends support to the hypothesis that mass transfer from a RSG to a companion results in the rapid evolution of the RSG to other stellar types~\citep{2008MNRAS.384.1109E}, such as Wolf--Rayet and helium stars~\citep{2001A&A...369..939W,2018A&A...615A..78G}. 
In addition to this channel, another plausible alternative evolutionary pathway for a RSG is the merger of two main-sequence stars.
This scenario results in a rejuvenated main-sequence star that will evolve toward the RSG evolutionary stage as a red straggler star (see B19).
From a RV variation point of view, these targets would be likely indistinguishable from a star that has evolved as a single star. 


\section{Conclusions} \label{sec:conclusion}

Estimated RVs are provided for a sample of 17 RSGs (including four lower luminosity AGB candidates) observed by the VFTS project. The targets are distributed across the 30 Doradus region of the LMC and include several stars in the Hodge\,301 and SL\,639 clusters. RVs are presented for each target from observations with three FLAMES--Giraffe settings. The RVs were estimated using a slicing technique, that is shown to provide accurate and precise results when compared with results for hot stars from the VFTS, as well as previous literature measurements for a subset of our targets. 
With the multi-epoch data we identify and correct for correlations in the measured RVs on the scale of $\pm$2\,\kms\/ for stars in all multi-object field configurations observed.
This correction was not possible for four of the target stars.

We provide estimates of the systemic velocities and dispersions of the
Hodge\,301 and SL\,639 clusters, which are in excellent agreement with results from B-type stars~\citep{2015A&A...574A..13E}. By combining the RSG velocities with those for the B-type stars we used the Virial theorem to estimate the dynamical masses of the two clusters.
For H\,301 we estimate $\log (M_{\rm dyn}/M_{\odot})=$~3.8\,$\pm$0.3, in excellent agreement with estimated mass from photometry~\citep{2016ApJ...833..154C}. We also obtained an upper limit  on the dynamical mass of SL\,639 of $\log (M_{\rm dyn}/M_{\odot})<$~3.1\,$\pm$\,0.8; this is the first mass estimate for SL\,639. 

The effectiveness of RSGs as kinematic tracers is discussed and we demonstrate that RSGs are robust tracers of the mean velocity and line-of-sight velocity dispersion of clusters given their lack of RV variability and apparent lack of binarity.
In the specific cases of H\,301 and SL\,639, the limiting factor on the measurements presented here, for solely RSGs, is the sample size.

By considering the multi-epoch data, we estimated an upper limit on the observed binary fraction of our sample. 
To do this, we first calculated the semi-amplitude velocities for RSG binary systems across a wide range of parameter space and found that the expected values are typically in the range 2\,$<$\,K\,$<$\,30\,\kms, for a 15\,M$_\odot$ RSG.
Using these calculations to guide our choice for the variability criteria used to detect binary candidates, we place an upper limit on the observed binary fraction of 0.3.
By simulating binary systems that can reproduce the observed data, we conclude that the binary fraction estimate is sensitive to systems within the parameter space q~$>$~0.3, $\log$P\,[days]~$<$~3.5, i.e. only among the shortest period systems allowed given the size of the RSG primary.

Using the empirically defined distribution of binary systems from~\cite{2017ApJS..230...15M}, we investigate the intrinsic binary fraction by simulating our observed data using a range of intrinsic binary fractions.
We found that our observations can be best reproduced by an intrinsic binary fraction of 0.3, accounting for observational biases, considering orbital periods $\log$P\,[days]~$<$~4.3.

To improve upon the binary statistics presented here, a more
comprehensive multi-epoch RV study (and preferably at greater spectral resolution) is required.

\begin{acknowledgements}
The authors would like to thank the referee, A. Tokovinin, for the careful review which has improved the article. The authors would like to thank T. Masseron for helpful discussions and providing the synthetic template used in this work, and to Jonathan Smoker and Liz Bartlett for discussions
regarding the wavelength stability of FLAMES. L.\,R.\,P, N.\,B and A.\,H. acknowledge support from grant AYA2017-010115 68012-C2- 1-P from the Spanish Ministry of Economy and Competitiveness (MINECO) and the Gobierno de Canarias for project ProID2017010115.
This work has made use of data from the European Space Agency (ESA) mission
{\it Gaia} (\url{https://www.cosmos.esa.int/gaia}), processed by the {\it Gaia}
Data Processing and Analysis Consortium (DPAC,
\url{https://www.cosmos.esa.int/web/gaia/dpac/consortium}). Funding for the DPAC
has been provided by national institutions, in particular the institutions
participating in the {\it Gaia} Multilateral Agreement.
This publication makes use of data products from the Two Micron All Sky Survey, which is a joint project of the University of Massachusetts and the Infrared Processing and Analysis Center/California Institute of Technology, funded by the National Aeronautics and Space Administration and the National Science Foundation.
MG acknowledges support from the European Research Council (ERC-StG-335936, CLUSTERS).
HS acknowledges support from the European Research Council (ERC-CoG-772225, MULTIPLES).
\end{acknowledgements}

\bibliographystyle{aa} 
\bibliography{journals}      

\begin{appendix}

\section{Multi-epoch radial velocities}
Table~\ref{tb:rvall} provides the complete list of estimated 
RVs for the candidate RSGs considered in this paper. Results for targets in Fields A, B, C and I are corrected for the apparent instrumental variations using at least two targets in their respective fields (see Section~\ref{sec:rv} for details).

\begin{table*}
\caption{Radial velocity (RV) estimates at each epoch for candidate RSGs observed in the VFTS. Observations with the same spectrograph setting for each target that were less than a day apart were coadded to improve signal-to-noise (for which the quoted epochs are the average of the coadded observations).}\label{tb:rvall}
\centering                                      

\begin{tabular}{lcccl}          
\hline\hline      
VFTS ID & HJD - 2\,400\,000 & RV (\kms) & $\sigma_{\rm RV}$ (\kms) & Setting \\

\hline
023 & 54767.824 & 271.30 & 0.61 & LR02 \\
023 & 54819.809 & 270.68 & 0.36 & HR15N \\
023 & 54827.772 & 271.88 & 1.41 & LR02 \\
023 & 54827.825 & 270.90 & 0.66 & HR15N \\
023 & 54828.751 & 273.49 & 0.84 & LR02 \\
023 & 54828.821 & 271.32 & 0.91 & LR03 \\
023 & 54836.563 & 269.57 & 1.32 & LR03 \\
023 & 54860.643 & 268.65 & 0.78 & LR02 \\
023 & 54886.616 & 268.59 & 0.97 & LR02 \\
023 & 55114.859 & 265.72 & 3.50 & LR02 \\
081 & 54804.637 & 286.16 & 0.62 & LR02 \\
081 & 54804.760 & 287.06 & 0.27 & HR15N \\
081 & 54808.676 & 285.06 & 0.90 & LR03 \\
081 & 54836.752 & 284.92 & 0.60 & LR02 \\
081 & 54867.598 & 283.11 & 0.57 & LR02 \\
081 & 55108.808 & 284.56 & 0.65 & LR02 \\
\ldots & \ldots & \ldots & \ldots & \ldots \\
\hline                                             
\end{tabular}
\tablefoot{The full version of this table is available electronically. The first 15 lines are shown here as a sample.
}
\end{table*}

\section{Semi-amplitude radial velocities for RSGs}
\label{ap:kcalc}

To investigate the expected range of semi-amplitude velocities ($K$) given the mass and physical extent of a RSG primary, we use the following equations,

\begin{equation}
  K = \frac{2\pi a_1}{P}
  \label{eq:k}
\end{equation}

where $P$ is the orbital period of the system and $a_1$ as,

\begin{equation}
  a_1 = \left(\frac{G m_2 P^2}{4\pi^2 (1 + q)^2}\right)^{1/3}
  \label{eq:a1}
\end{equation}

where $q = m_1/m_2$ and $m_1$, $m_2$ are the mass of the primary and secondary components of the system, respectively. 

\begin{figure}[btp]
  \centering
  \includegraphics[width=\columnwidth]{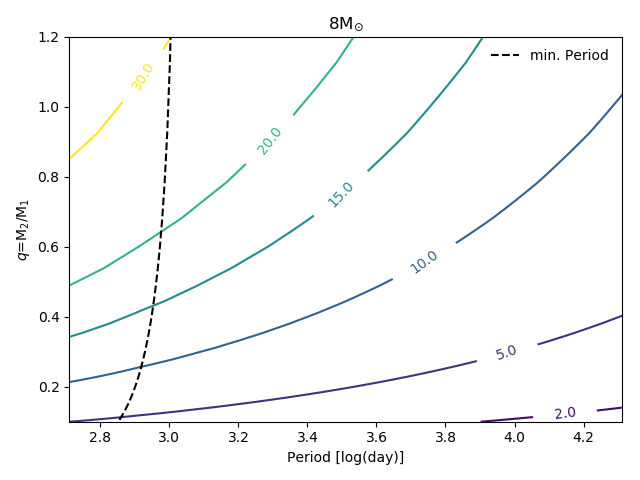}  
  \caption[]{\label{fig:qvsP8} %
  Simulated semi-amplitude ($K$) velocities for an 8\,M$_{\odot}$ RSG in a binary system for a range of periods and mass ratios. 
  Coloured solid lines illustrate constant velocities at 2, 5, 10, 15, 20, 30\,\kms, from shallow to steep respectively. 
  The black dashed line highlights the minimum possible period for an 8\,M$_{\odot}$ RSG for the corresponding mass ratio assuming that its maximum radius is 400\,R$_{\odot}$~\citep{2011A&A...530A.115B}. Therefore, all systems to the left of this line are excluded as a result of the size of the RSG primary.
  }
\end{figure}

\begin{figure}[btp]
  \centering
  \includegraphics[width=\columnwidth]{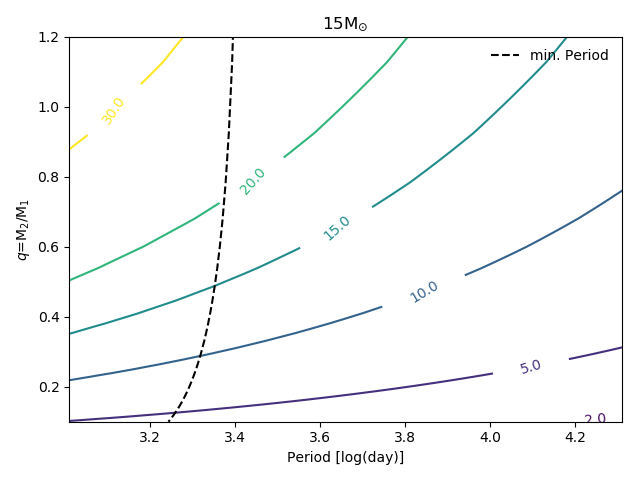}  
  \caption[]{\label{fig:qvsP15} %
  As for Figure~\ref{fig:qvsP8} for a 15\,M$_\odot$ RSG primary, where the corresponding maximum radius is 900\,R$_{\odot}$~\citep{2011A&A...530A.115B}.
  }
\end{figure}

Figures~\ref{fig:qvsP8} and~\ref{fig:qvsP15} illustrate the cases for 8 and 15\,M$_{\odot}$ RSGs, where the solid coloured lines illustrate constant $K$ values and the black dashed lines represent the minimum period possible, given the maximum radii of RSGs, for different mass ratios. 
In the range $q$\,$>$\,0.1, 4.5\,$<$\,P [year]~$<$\,50, we expect 2\,$<$\,K\,$<$\,30\,\kms\ for a 15\,M$_\odot$ RSG.

Orbital eccentricity and inclination are not taken into account in these calculations but would both act to decrease the observed semi-amplitude velocity.

\end{appendix}

\end{document}